\documentclass[%
reprint,
onecolumn,
 amsmath,amssymb,
 aps,
]{revtex4-2}

\usepackage{graphicx}
\usepackage[hidelinks]{hyperref}

\usepackage{dcolumn}
\usepackage{bm}
\usepackage{amsmath}
\usepackage{mathtools}
\usepackage{amssymb}
\usepackage{makecell}
\usepackage{color}
\usepackage{ulem}
\usepackage{enumitem}

\usepackage{comment}

\newcommand{\mdeg}{\mathcal{D}}
\newcommand{\md}{\mathcal{M}}

\begin{document}

\title{Communities for the Lagrangian Dynamics of the Turbulent Velocity Gradient Tensor: A Network Participation Approach}

\author{Christopher ~J.~Keylock${}^{1,2}$}
\author{Maurizio Carbone${}^{2,3,4}$}
\email{C.J.Keylock@lboro.ac.uk}

\affiliation{${}^{1}$Loughborough University, School of Architecture, Building and Civil Engineering, Loughborough, Leicestershire, U.K.\\
${}^{2}$University of Bayreuth, Department of Theoretical Physics, Bayreuth, 95440, Germany\\
${}^{3}$Istituto dei Sistemi Complessi, CNR, Via dei Taurini 19, 00185, Rome, Italy\\
${}^{4}$INFN “Tor Vergata”, Via della Ricerca Scientifica 1, 00133, Rome, Italy}

\date{\today}

\begin{abstract}
Complex network analysis methods have been widely applied to nonlinear systems, but applications within fluid mechanics are relatively few. In this paper, we use a network for the Lagrangian dynamics of the velocity gradient tensor (VGT), where each node is a flow state, and the probability of transitioning between states follows from a direct numerical simulation of statistically steady and isotropic turbulence. The network representation of the VGT dynamics is much more compact than the continuous, joint distribution of a set of invariants for the tensor. We focus on choosing optimal variables to discretize and classify the VGT states. To this end, we test several classifications based on topology and various properties of the background flow coherent structures. We do this using the notion of ``community'' or ``module'', namely clusters of nodes that are optimally distinct while also containing diverse nodal functions. The best classification based upon VGT invariants often adopted in the literature combines the sign of the principal invariants, $Q$ and $R$, and the sign of the discriminant function, $\Delta$, separating regions where the VGT eigenvalues are real and complex. We further improve this classification by including the relative magnitude of the non-normal contribution to the dynamics of the enstrophy and straining stemming from a Schur decomposition of the VGT. The traditional focus on the second VGT principal invariant, $Q$, implies consideration of the difference between the enstrophy and strain-rate magnitude without the non-normal parts. The fact that including the non-normality leads to a better VGT classification highlights the importance of unclosed and complex terms contributing to the VGT dynamics, namely the pressure Hessian and viscous terms, to which the VGT non-normality is intrinsically related.
\end{abstract}

\maketitle











\section{Introduction}
\label{introduction}
The use of symbolic dynamics to characterize fluid flow dates back to \cite{hadamard98} who studied geodesics of surfaces of negative curvature in terms of a sequence of symbols. Such an approach has been used relatively sparingly since, with a few theoretical contributions \cite{bowen73}, and applications in e.g., geosciences, where it helped characterize the interaction between large-scale flow structures and natural, mobile boundaries from single-point velocity measurements \cite{kirkbride95,k14}. 
More generally, in nonlinear time-series analysis, symbolic sequence analysis has been employed to enhance the signal-to-noise ratio \cite{beimgraben01} and to underpin the construction of discrete, complex network approaches to model continuous systems \cite{mccullough15} with applications in a range of fields, such as neuroscience \cite{telesford11}, transport planning \cite{wang11} and electrical power provision \cite{crucitti04}. In fluid dynamics, complex network approaches have been relatively under-utilized. Examples include the investigation of vortex interaction in two-dimensional turbulence \cite{nair15,taira16} and the study of time irreversibility in near-wall flows \cite{Iacobello2023}. However, complex networks have typically not been used to study the dynamics of the turbulent velocity gradient tensor (VGT) that is the focus of this work.

An open question in the study of the VGT is what is the best way to classify the flow dynamics into different communities of flow states? Such a question has typically been considered from a continuum perspective, rather than utilising discrete networks, and has typically been based upon defining communities of flow structures based on topological properties of the VGT \cite{PC87}, although there have been noteworthy departures from the standard approaches \cite{luthi09}. Classifying the VGT states in terms of its topology is a question that is closely related to the definitions of coherent flow structure in turbulent flows \cite{hunt88,dubief00,chakraborty05,xu19}.

Examples of such classifications in nonlinear physics include typical investigations into cliques \cite{bollobas76}, communities \cite{fortunato10}, or modules \cite{guimera05}. Those entail applying some form of partition to extract a set of discrete and non-overlapping ``units'' that, when taken together include all nodes in a network (although the definition of clique due to \cite{palla05} permits these to be nested within a given module or overlap with more than one module). In general, modules provide information about the network structure at an intermediate scale between individual nodes (e.g.~measures of node centrality \cite{bavelas50,freeman77}) and the whole network (such as contrasting the exponential or power-law degree distributions of, respectively, the Erd{\"{o}s}-R{\'{e}}nyi \cite{ER59} and Barab{\'{a}}si and Albert \cite{BA99} networks). In this respect, eigenvector centrality \cite{bon72} has an interesting status as it is node-oriented while also encompassing information on how each node is connected to other significant nodes in the network. 

While in complex physics a search for communities in a network is typically empirical, in fluid dynamics, we can formulate communities \textit{a priori}, based on quantities classically considered in the literature, and then test their effectiveness. This is the approach adopted in this paper: we first provide a background on the relevant invariants and properties of the VGT, then use them to define communities in a complex network representing the VGT dynamics and finally classify the resulting communities. We build the complex network and analyze the communities using Lagrangian data from a direct numerical simulation of homogeneous, isotropic turbulence (HIT) \cite{yili}.


\section{The Dynamics of the Velocity Gradient Tensor}

Our starting point is the continuity equation and Navier-Stokes momentum equation for an incompressible, Newtonian fluid: 
\begin{align}
\nabla \cdot \bm{u} &= 0, \\
\frac{\partial \bm{u}}{\partial t} + (\bm{u} \cdot \nabla)\bm{u} &= -\nabla p + \nu \nabla^2 \bm{u},
\label{eq.NS}
\end{align}
where $\bm{u}(\bm{x},t)$ is the three-dimensional velocity vector field, $t$ is time, $p(\bm{x},t)$ is the pressure field divided by the constant density, and $\nu$ is the kinematic viscosity. The VGT is defined as $\bm{A}\equiv\nabla\bm{u}$ and its evolution equation is obtained by taking the spatial gradient of (\ref{eq.NS}):
\begin{equation}
\frac{\partial \bm{A}}{\partial t} + (\bm{u} \cdot \nabla)\bm{A} = -\left(\bm{A}^2-\frac{\mbox{tr}(\bm{A}^2)}{3}\bm{I}\right) - \bm{H} + \nu \nabla^2 \bm{A},
\label{eq.Aevolve}
\end{equation}
where $\bm{I}$ is the identity matrix, $\mbox{tr}(\ldots)$ indicates the matrix trace, and $\bm{H}$ is the anisotropic pressure Hessian, $\bm{H}\equiv \bm{\nabla \nabla} p -2Q\bm{I}/3$, with $Q\equiv -\mbox{tr}(\bm{A}^2)/2$ the second principal invariant of the VGT. 

The relation between the trace of the pressure Hessian and the second principal invariants of the VGT highlights the relevance of the VGT invariants, because of their dynamical relevance \cite{betchov56,vieillefosse82}, their role in parameterizing the single-point statistics of the VGT is isotropic turbulence \cite{Itskov2015,Carbone2023} and importance for the classification of small-scale flow states \cite{Sharma2021}.
Following \cite{cantwell93}, we can use (\ref{eq.Aevolve}) to write down equations for the Lagrangian evolution of the VGT principal invariants, $Q$ and $R\equiv -\mbox{tr}(\bm{A}^3)/3$, in terms of each other, formally including the viscous effects and the coupling between the VGT $\bm{A}$, the deviatoric pressure Hessian $\bm{H}$ and viscous contributions $\nu\nabla^2\bm{A}$:
\begin{align}
\label{eq.evolveQ}
\frac{d Q}{d t} &= -3R - \nu\,\,\mbox{tr}\left(\bm{A} \nabla^{2}\bm{A}\right) + \mbox{tr}\left( \bm{A} \bm{H}\right) \\
\label{eq.evolveR}
\frac{d R}{d t} &= \frac{2}{3}Q^{2} - \nu\,\,\mbox{tr}\left(\bm{A}^{2} \nabla^{2}\bm{A}\right) + \mbox{tr}\left( \bm{A}^{2} \bm{H} \right),
\end{align}
where $d/dt$ is the Lagrangian derivative taken along fluid particle trajectories. 
Neglecting the last two terms on the right-hand side of (\ref{eq.evolveQ}) and (\ref{eq.evolveR}) yields the restricted Euler model \cite{vieillefosse82,cantwell92}, in which $Q$ and $R$ and thus the VGT eigenvalues $\lambda_i$ are given by the characteristic equation
\begin{equation}
\lambda_{i}^{3} + P \lambda_{i}^{2} + Q \lambda_{i} + R = 0,
\label{eq.inv}
\end{equation}
where the first invariant, $P$, is zero in incompressible flow. The principal invariants, $Q$ and $R$ 
evolve independently of the other VGT components but the restricted Euler model features a finite-time blowup for almost all initial conditions, which is not observed in numerical simulations of the Navier-Stokes equations. This indicates that the deviatoric pressure Hessian and viscous terms crucially enter the VGT dynamics by affecting the VGT magnitude and alignments. In particular, the effects of the pressure and viscous terms reflect on the statistics of the strain-rate tensor $\bm{S}$ and rotation-rate tensor $\bm{\Omega}$, 
\begin{align}
\bm{S} &= \frac{1}{2}(\bm{A} + \bm{A}^{*}) \\
\bm{\Omega} &= \frac{1}{2}(\bm{A} - \bm{A}^{*})
\end{align}
together with their statistical alignments \cite{tsinober97}.
Formulating models for the dynamics that correctly capture the complexities introduced by the non-local effects of the pressure Hessian has been the focus of a significant body of work, e.g.~\cite{girimaji90,martin98,chevillard08,wilczek14,johnson16,Buaria2023,Das2024,Carbone2024}, as reviewed in \cite{meneveau11,wilczek24}.

Here, we explicitly separate the normal part of $\bm{A}$, related to its eigenvalues $\lambda_i$, and the non-normal part of $\bm{A}$ by using a complex Schur transform \cite{schur09}. This transform imposes a unitary form for the rotation vectors and moves the non-normality, $\bm{N}$, into the upper-triangular part of the central matrix of the transform, $\bm{T}$, \cite{k18}
\begin{equation}
\label{eq.schur}
    \bm{A} = \bm{U}\bm{T}\bm{U}^{*},
\end{equation}
where $\bm{T} = \bm{\Lambda} + \bm{N}$, $\bm{\Lambda}$ is a diagonal matrix of eigenvalues such that $\Lambda_{i,i} = \lambda_{i}$, and $\bm{N}$ is upper-diagonal. The normal and non-normal contributions to the dynamics may be isolated according to, $\bm{B} = \bm{U}\bm{L}\bm{U}^{*}$ and $\bm{C} = \bm{U}\bm{N}\bm{U}^{*}$. The principal invariants $Q$ and $R$ can be written purely in terms of strain and rotation components of $\bm{B}$, while the enstrophy, total straining and production terms all involve contributions from $\bm{C}$ \cite{k18}. Hence, we have
\begin{subequations}
\begin{align}
\label{eq.Q}
Q &= \frac{1}{2}\left(\Vert\bm{\Omega} \Vert^{2} - \Vert\bm{S} \Vert^{2}\right),\\
\Vert\bm{\Omega}\Vert^{2} &= \Vert\bm{\Omega}_{B}\Vert^{2} + \Vert\bm{\Omega}_{C}\Vert^{2},\\
\Vert\bm{S}\Vert^{2} &= \Vert\bm{S}_{B}\Vert^{2} + \Vert\bm{\Omega}_{C}\Vert^{2},
\end{align}
\label{eq.Qterms}
\end{subequations}
where $\Vert\ldots\Vert$ is the Frobenius norm. The equivalent expressions for the third invariant are
\begin{subequations}
\begin{align}
\label{eq.R}
R &= -\mbox{det}\left(\bm{S}\right) - \mbox{tr}\left(\bm{\Omega}^{2}\bm{S} \right),\\
-\mbox{det}\left(\bm{S}\right) &= -\mbox{det}\left(\bm{S}_{B}\right) + \mbox{tr}\left(\bm{\Omega}_{C}^{2}\bm{S}_{B}\right) -\mbox{det}\left(\bm{S}_{C}\right),\\
\mbox{tr}\left(\bm{\Omega}^{2}\bm{S}\right) &= \mbox{tr}\left(\bm{\Omega}_{B}^{2}\bm{S}_{B}\right) + \mbox{tr}\left(\bm{\Omega}_{C}^{2}\bm{S}_{B}\right) -\mbox{det}\left(\bm{S}_{C}\right),
\end{align}
\label{eq.Rterms}
\end{subequations}
where $\mbox{det}(\ldots)$ is the determinant. This decomposition has been adopted to gain an insight into several properties of turbulence including the reason for the preferred alignment between the vorticity vector and the eigenvector corresponding to the intermediate eigenvalue of the strain rate tensor \cite{k18}, and the the physics of the flow when non-normality is maximal \cite{k19}. In addition, in spatially developing flows, it has been shown that $\Vert\bm{\Omega}_{C}\Vert^{2}$ is crucial for the dynamics before turbulence is fully established (i.e.~regions where pressure Hessian contributions typically dominate the kinetic energy budget) \cite{beaumard19}, 
and to characterize the role of in-rushing sweeps in boundary layers \cite{bt86,k22}. 

In this study, we make use of the terms in (\ref{eq.Qterms}) and (\ref{eq.Rterms}) to provide a means to expand the possible number of physical quantities that may be relevant for effective classification of the VGT dynamics. Hence, we form a network where the nodal attributes consist of the signs of these terms as well as the relative rankings of their magnitudes. That is, the rank-order of the three terms on the right-hand side of (\ref{eq.Qterms}) all of which are non-negative, and then the rank-order of the absolute values of the terms on the right-hand side of (\ref{eq.Rterms}): $\vert\mbox{det}\left(\bm{S}_{B}\right)\vert$, $\vert\mbox{tr}\left(\bm{\Omega}_{B}^{2}\bm{S}_{B}\right)\vert$, $\vert\mbox{tr}\left(\bm{\Omega}_{C}^{2}\bm{S}_{B}\right)\vert$ and $\vert\mbox{det}\left(\bm{S}_{C}\right)\vert$. We refer to \cite{k24} for the definition of the network nodes.
Given such a network, and with different classifications making use of different combinations of terms as explained in the next section, we can then make use of network community extraction techniques to determine the best classification for the Lagrangian dynamics of the VGT.  

\section{\textit{A priori} Network Classifications}

\begin{table}
\begin{tabular}{l l l} 
 \hline
 Sign($\Delta$) & Sign($R$) & Topology \\ 
 \hline
-  & - & stable-node/saddle/saddle (SN/S/S) \\
-  & + & unstable-node/saddle/
saddle (UN/S/S) \\
+ & - & stable-focus/stretching (SF/S)\\
+ & + & unstable-focus/
contracting (UF/C)\\
\hline
\end{tabular}
\caption{Topological flow states in the $Q$-$R$ plane following \cite{ooi99}.
}
\label{table.perry}
\end{table}

Rather than adopting a purely empirical approach to community network classification, our approach is to test existing, physics-based classifications of the VGT dynamics, as well as extensions of them based on the decompositions given in (\ref{eq.Qterms}) and (\ref{eq.Rterms}). In this section, we outline the basis for the ten primary classifications examined in this paper.

From the restricted Euler model given by the simplification of (\ref{eq.evolveQ}) and (\ref{eq.evolveR}) by disregarding the last two terms on the right-hand side of each of these equations, it follows that the dynamics on the $Q$-$R$ plane \cite{vieillefosse82,vieillefosse84} are the logical starting point for looking to define communities. This approach increased in popularity significantly following the work of \cite{PC87} and \cite{ooi99} (see \cite{meneveau11} for a review), as it provided a topological classification of four different flow configurations based on the sign of $R$ and the sign of the discriminant function, $\Delta$, 
\begin{equation}
\label{eq.Delta}
\Delta = Q^{3} + \frac{27}{4}R^{2},
\end{equation}
which separates the regions based on whether the eigenvalues for $\mathbf{A}$ are real or complex.
These four topologies are defined in Table \ref{table.perry} and while this provides a means to classify the flow into different, non-overlapping modules/communities, it is not the only approach possible on the $Q$-$R$ plane. For example, a positive value of $Q$ has been used extensively as a criterion for coherent flow structure identification \cite{hunt88,dubief00}, which follows from the physical interpretation of $Q > 0$ as the excess enstrophy (\ref{eq.Q}). Hence, an alternative definition of modules/communities to that in Table \ref{table.perry} would be one based on the signs of $Q$ and $R$. Thus, we have:
\begin{enumerate}[label=\textbf{C\arabic*},leftmargin=*]
\item A classification into $N_{M} = 4$ modules based on the sign of $R$ and the sign of the discriminant function, $\Delta$ \cite{PC87};
\item A classification into $N_{M} = 4$ modules based on the sign of $R$ and the sign of $Q$.
\end{enumerate}

A superposition of \textbf{C1} and \textbf{C2} leads to a classification of the VGT in terms of six regions and it has been shown that there are distinct behaviours in each of these regions even though they are not all topologically distinct \cite{k18}. Thus, our third classification is a hybrid of the first two:
\begin{enumerate}[label=\textbf{C\arabic*},leftmargin=*]
\setcounter{enumi}{2}
\item A classification into $N_{M} = 6$ modules based on the signs of $R$, $Q$ and $\Delta$. 
\end{enumerate}

A departure from working in the $Q$-$R$ plane was introduced by \cite{luthi09} who separated $R$ into the terms on the right-hand side of (\ref{eq.R}). This leads to
\begin{enumerate}[label=\textbf{C\arabic*},leftmargin=*]
\setcounter{enumi}{3}
\item A classification into $N_{M} = 8$ modules based on the sign of $Q$ and then the signs of the strain production, $-\mbox{det}(\bm{S})$ and the enstrophy production, $\mbox{tr}(\bm{\Omega}^{2}\bm{S})$.
\end{enumerate}

Considering our expansion of $R$ in (\ref{eq.Rterms}) it is clear that \textbf{C4} permits the combined effect of the non-normal production $\mbox{det}\left(\bm{S}_{C}\right)$ and the interaction production $\mbox{tr}\left(\bm{\Omega}_{C}^{2}\bm{S}_{B}\right)$ to be included in analysis (while these terms cancel when considering $R$). However, because both of these terms feature in both equations in (\ref{eq.Rterms}) there is an innate correlation between these variables. While the signs of the normal strain production,  $\mbox{det}\left(\bm{S}_{B}\right)$ and the normal enstrophy production, $\mbox{tr}\left(\bm{\Omega}_{B}^{2}\bm{S}_{B}\right)$ are necessarily opposite where $\Delta > 0$, (with $\mbox{sgn}[-\mbox{det}(\mathbf{S}_{B})] = \mbox{sgn}[R]$), there is no similar constraint on the signs for the other two terms. Thus, one may extend \textbf{C4} to form
\begin{enumerate}[label=\textbf{C\arabic*},leftmargin=*]
\setcounter{enumi}{4}
\item A classification with $N_{M} = 16$ modules based on the signs of $Q$ and $R$, with the latter setting the signs for $-\mbox{det}(\mathbf{S}_{B})$ and $\mbox{tr}(\mathbf{\Omega}_{B}^{2}\mathbf{S}_{B})$, and then the signs of $-\mbox{det}(\mathbf{S}_{C})$ and $\mbox{tr}(\mathbf{\Omega}_{C}^{2}\mathbf{S}_{B})$.  
\end{enumerate}

Our final general classification type extends the logic \cite{luthi09} applied to $R$ to also consider $Q$. From (\ref{eq.Qterms}) we have the non-normality, $\Vert\bm{\Omega}_{C}\Vert^{2}$, in addition to the normal enstrophy, $\Vert\bm{\Omega}_{B}\Vert^{2}$ and the normal straining, $\Vert\bm{S}_{B}\Vert^{2}$. As all these terms are non-negative, it only makes sense to consider the sign of their differences as in (\ref{eq.Q}). Hence, with $Q$ retained to contrast the magnitudes of the normal enstrophy and normal strain, we may define
\begin{align}
\chi &= \Vert\boldsymbol{\Omega}_{C}\Vert^{2} - \Vert\boldsymbol{\Omega}_{B}\Vert^{2}\\
\xi &= \Vert\boldsymbol{\Omega}_{C}\Vert^{2} - \Vert\mathbf{S}_{B}\Vert^{2},
\label{eq.xichi}
\end{align}
and there will be six possible values for the signs of $Q$, $\chi$ and $\xi$ when taken together (because if $Q > 0$ and $\chi > 0$, necessarily $\xi > 0$; likewise if $Q < 0$ and $\xi > 0$, necessarily $\chi > 0$). If, in addition, we follow \textbf{C3} and incorporate the sign of $\Delta$ into our analysis then there is a further constraint that restricts the number of combinations of signs that can be observed because $\Vert\boldsymbol{\Omega}_{B}\Vert^{2} = 0$ when $\Delta < 0$, which imposes $\chi < 0$. Hence, our sixth classification type is
\begin{enumerate}[label=\textbf{C\arabic*},leftmargin=*]
\setcounter{enumi}{5}
\item A classification with $N_{M} = 64$ modules based on the signs of $Q$, $\chi$, $\xi$, $\Delta$, $R$, $-\mbox{det}(\mathbf{S}_{C})$ and $\mbox{tr}(\mathbf{\Omega}_{C}^{2}\mathbf{S}_{B})$.  
\end{enumerate}

The introduction of the variables $\xi$ and $\chi$ means that we may also extend the classifications \textbf{C2} to \textbf{C5} (all of which involve $Q$) by introducing these two additional variables, which increases the number of modules by a factor of 3 where $\Delta > 0$ and by a factor of 2 where $\Delta < 0$. Thus, in addition to \textbf{C1} to \textbf{C6} we have four additional categories, \textbf{C2b}-\textbf{C5b}, which extend \textbf{C2}-\textbf{C5} by including $\xi$ and $\chi$.

\section{Network modularity and participation}
\subsection{Construction of our network}
A network or graph, $\mathcal{G}$, without self-loops, comprises a set of $N$ nodes (vertices), $\mathcal{V}$, edges $\mathcal{E}$, and weights, $\mathcal{W}$. We construct a network with $N = 844$ vertices based on the 64 states defined for \textbf{C6} as well as the relative ranking of the magnitudes of the four production terms in (\ref{eq.Rterms}) \cite{k24}. Thus, we have approximately thirteen times as many nodes as there are possible modules/communities in classification \textbf{C6}. The adjacency matrix, $\bm{W}$, consists of zeros along the primary diagonal (because there are no self-loops) and then weights, $w_{ij}$ based on the probability of transitioning from node $i$ to $j$, i.e.~the flow changing from a state described by one node to a state described by another. These transition probabilities were determined based on Lagrangian tracking of $27^{3}$ tracers in a direct numerical simulation of homogeneous isotropic turbulence (HIT) \cite{yili} with an initial separation of two Taylor microscales. Particles were followed for 240 Kolmogorov times, $\tau_{\eta}$ at a resolution of $0.05 \tau_{eta}$. The ten network classifications were then applied to the 844-node network to determine the optimal partitioning of the flow states. Further details on the network construction are provided in \cite{k24}.  

\subsection{Modularity}
The notion of a community or ``module'' within a network can be formalized to enable a quantitative analysis of intermediate scale structure between that of the individual vertex and the network as a whole \cite{guimera05}. An important principle to find optimal partitions has been to maximize the modularity \cite{newman04}, which for a given partition into $N_{M}$ distinct modules is given by:
\begin{equation}
\label{eq.module}
M \equiv \sum_{k=1}^{N_{M}} \left [ \frac{S_{k}}{S_{W}}-\frac{\mdeg_{k}}{2S_{W}}   \right ].
\end{equation}
Here $N_{M}$ is the number of modules, 
\begin{equation}
S_{W} = \sum_{i=1}^{N}\sum_{j=1}^{N} W_{ij}
\end{equation}
is the sum of all weights in the network, 
$S_{k}$ is the sum of all weights over all nodes in module $k$
\begin{equation}
   S_{k} = \sum_{i \in \md_k}\sum_{j \in \md_k} W_{ij},
\end{equation}
where $\md_k$ is the set of all node indices in module $k$, and
\begin{equation}
\label{eq.moddeg}
   \mdeg_{k} = \sum_{i \in \md_k} D_{i},
\end{equation}
is the module degree, given by the sum of the degrees $D_i=\sum_{j=1}^{N} W_{ij}$ of all the nodes in module $k$. We normalized $\mathbf{W}$ such that $S_{W} = \sum_{k=1}^{N_{M}} S_{k} = 1$.
The nature of (\ref{eq.module}) is such that the second term within the bracket prevents a maximization of $S_{k}$ that leads to a single module corresponding to the whole network. In this paper, we adopted an alternative approach to this maximization problem. Noting that $p_{M}(k) = S_{k}/S_{W}$ is in the form of a probability, we adopted an entropy approach, by employing the Boltzmann-Gibbs-Shannon entropy
\begin{equation}
\mathcal{E} = -\sum_{k=1}^{N_{M}} p_{M}(k) \log_{2} p_{M}(k).
\label{eq.entropy}
\end{equation}
This formalism intrinsically prevents an optimum corresponding to the whole network because $N_{M} = 1$ gives $\mathcal{E} = 0$. While it is the case that \emph{ceteris paribus} $\mathcal{E}$ grows with $N_{M}$ it will increase less steeply if further division into new modules does not result in significant entropy increase. The entropy approach also has the useful property that for a given $N_{M}$, $\mathcal{E}$ is maximized when the total edge probabilities are equally distributed over all modules.   

\subsection{Participation}
Optimizing the modularity does not distinguish between different forms of nodal function and the extent to which these are represented within a given module. This can be investigated using the participation coefficient, which for node $i$ is defined as \cite{guimera05}
\begin{equation}
 \rho_{i} = 1 - \sum_{k=1}^{N_{M}} \left ( \frac{\mdeg_{ik}}{D_{i}} \right )^{2},
 \label{eq.part}
\end{equation}
where 
\begin{equation}
   \mdeg_{ik} = \sum_{j \in \md_k} W_{ij}.
\end{equation}
Hence, if a vertex is only connected to other nodes in its own module, $i \in \md_k$, then $\rho_{i} = 0$, while even connectivity to all modules gives $\rho_{i} \to 1$.
Again, a probabilistic interpretation of $\rho_{i}$ is possible if we define $p_{D}(i,k) = \mdeg_{ik} / D_{i}$. As noted by \cite{cajic24} this means that (\ref{eq.part}) may be interpreted as equivalent to a Gini coefficient, which is derived from the Shannon entropy by expanding the logarithm and truncating to the leading term. 

It was proposed by \cite{guimera05} to complement $\rho_{i}$ with a metric for within-module connectivity based on the statistical notion of a $z$-score. With $\mdeg_{ik}^{+}$ the variant of $\mdeg_{ik}$ that is restricted/conditioned to nodes in the same module,
i.e.~$\mdeg_{ik}^{+} = \sum_{j \in \md_k} \left( W_{ij} \mid i \in \md_k\right)$, then   
\begin{equation}
\label{eq.z}
z_{i} = \frac{\mdeg_{ik}^{+} - \overline{\mdeg}_{ik}^{+}}{\sigma_{D}(k)},
\end{equation}
where the overbar indicates the mean over all nodes in module $k$ and the standard deviation of this set of nodes is given by $\sigma_{D}(k)$.

\begin{table}
\begin{tabular}{l l l} 
 \hline
 Role & Hub/non-hub & Description \\ 
 \hline
Ultra-peripheral  & non-hub & $\rho_{i} \sim 0$ \\
Peripheral & non-hub & $\rho_{i} \le 0.625$\\
Connector & non-hub & $0.625 < \rho_{i} \le 0.8$\\
Kinless  & non-hub & $\rho_{i} > 0.8$\\
Provincial & hub &  Node with a large degree\\ & & and $\rho_{i} \sim 0.3$\\
Connector & hub & Node with a large degree\\ & & and $0.3 < \rho_{i} \le 0.75$\\
Kinless & hub & $\rho_{I} > 0.75$\\
 \hline
\end{tabular}
\caption{Classification of node function based on $z_{i}$ and $\rho_{i}$ following \cite{guimera05} A non-hub has $z < 2.5$ while $z_{i} \ge 2.5$ corresponds to a hub.
}
\label{table.guimera}
\end{table}

Based on a threshold applied to $z$ it was then suggested that $z_{i} \ge 2.5$  was a \emph{hub} node and $z_{i} < 2.5$ was a non-hub. Combining this threshold with $\rho_{i}$ resulted in the seven-state classification given in Table \ref{table.guimera}. 

\section{Analysis} 
\begin{figure}
\centering 
\includegraphics[width=0.8\textwidth, angle=0]{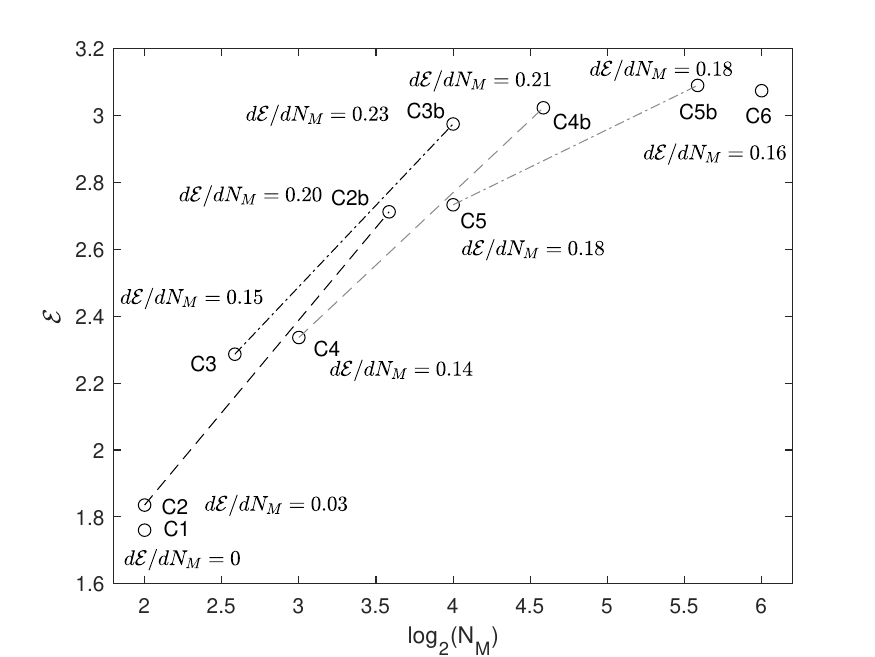}	
\caption{Values of the modularity entropy, $\mathcal{E}$ as a function of $N_{M}$ for each of the ten classifications. The lines connecting points illustrate the entropy change from introducing $\chi$ and $\xi$. Values for the semi-normalized gradient (\ref{eq.grad}) are stated next to each point.} 
\label{fig.entropy}
\end{figure}

Fig.~\ref{fig.entropy} shows the entropy, $\mathcal{E}$ from (\ref{eq.entropy}) as a function of $N_{M}$, which is displayed on a base-2 logarithmic scale as many of our classifications lead to $N_{M}$ being an integer power of two. What is clear from this maximization perspective is that $\mathcal{E}$ saturates at large $N_{M}$ when the number of vertices per module becomes small. Indeed, $\mathcal{E}$ for \textbf{C6} is lower than for \textbf{C5b} despite the greater value of $N_{M}$ for the former. Direct comparisons for the same $N_{M}$ show that \textbf{C2} is a better classifier than \textbf{C1}, (i.e.~$Q$ is a more effective variable than $\Delta$) and that \textbf{C3b} is a better classifier than \textbf{C5} (i.e., it is more important to include the signs of $\chi$ and $\xi$ than the signs of $-\mbox{det}(\mathbf{S}_{C})$ and $\mbox{tr}(\mathbf{\Omega}_{C}^{2}\mathbf{S}_{B}$).
The four lines added to the plot illustrate the impact of adding $\chi$ and $\xi$ to \textbf{C2}-\textbf{C5} and the degree of improvement is very similar for \textbf{C3} and \textbf{C4}, is greatest for \textbf{C2} to \textbf{C2b} and smallest for \textbf{C5} to \textbf{C5b}, where $\mathcal{E}$ approaches its maximum.
Noting that \textbf{C1} has the smallest value of $\mathcal{E}$ and \textbf{C5b} the largest, we can define a semi-normalized gradient as
\begin{equation}
\frac{d \mathcal{E}}{d N_{M}} = \frac{(\mathcal{E} - \mathcal{E}_{C1}) / (\mathcal{E}_{C5b} - \mathcal{E}_{C1})}{\log_{2}(N_{M})},
\label{eq.grad}
\end{equation}
and as shown by the labels in Fig.~\ref{fig.entropy}, this is maximal for \textbf{C3b} (followed by \textbf{C4b} and then \textbf{C2b}). Hence, the favoured classification is one with six regions based on the signs of $Q$, $R$ and $\Delta$, with the addition of $\chi$ and $\xi$ to incorporate the relative magnitude of $\Vert\bm{\Omega}_{C}\Vert^{2}$.

\begin{table}
\begin{tabular}{l r r r} 
 \hline
 Type & No. of nodes & $\langle \log_{10}(D_{i}^{O}) \rangle$ & $\sigma\left[\log_{10}(D_{i}^{O})\right]$\\ 
 \hline
Stable hub & 44 & -2.17 & 0.26\\
Non-stable hub & 110 & -2.70 & 0.25\\
Stable non-hub & 346 & -3.82 & 0.44\\
Semi-stable 1 non-hub & 52 & -3.43 & 0.30\\
Semi-stable 2 non-hub & 55 & -3.29 & 0.34\\
Non-stable non-hub & 234 & -3.17 & 0.32 \\
\hline
& 841 & &\\
 \hline
 All nodes & 844 & -3.34 & 0.60\\
 \hline
\end{tabular}
\caption{Node types and their stability based on their probability of changing function over the different classifications, as explained in the text, and the associated properties of $D_{i}^{O}$ for each type.
}
\label{table.ntypes}
\end{table}

Focusing on the individual vertices, Table \ref{table.ntypes} classifies the network nodes into hub or non-hub classes and then further sub-divides them based on how consistent this was over the ten classifications. The mean and standard deviation of the out-degree is also stated for each case (the out-degree is approximately normally distributed after a logarithmic transformation). Three nodes were classified as hub or non-hub an equal number of times and therefore excluded from this table.
Three hundred and ninety vertices did not alter their classification from hub or non-hub over all ten classifications. These are denoted as ``stable'' in Table \ref{table.ntypes}. Semi-stable type 1 nodes were those that were the same for all classifiers except for one of the two with the greatest number of modules (either \textbf{C5b} or \textbf{C6}). Semi-stable type 2 nodes were those that were the same for all classifiers except both \textbf{C5b} and \textbf{C6}; all of these cases were non-hub nodes. Otherwise, the non-stable cases were classified as hubs or non-hubs the majority of times, but the classifications that gave alternate results were not restricted to the high $N_{M}$ cases of \textbf{C5b} and \textbf{C6}.

Based on a \emph{t}-test, all of the means quoted in the table are significantly different at the 1\% significance level except those for the two semi-stable non-hub node groupings ($p = 0.03$) as might be expected given the similar definition of these two groupings. As there is close to a factor of 50 difference in the mean out-degree for the stable hubs compared to the stable non-hubs and given that a large degree forms part of the description of a hub in Table \ref{table.guimera}, then Table \ref{table.ntypes} indicates that 
we can robustly discriminate between the stable hubs and stable non-hubs. Hence, we can use the $\rho_{i}$ values from Table \ref{table.guimera} to classify these hubs and non-hubs into their different functions.  

\begin{figure}
\centering 
\includegraphics[width=0.8\textwidth, angle=0]{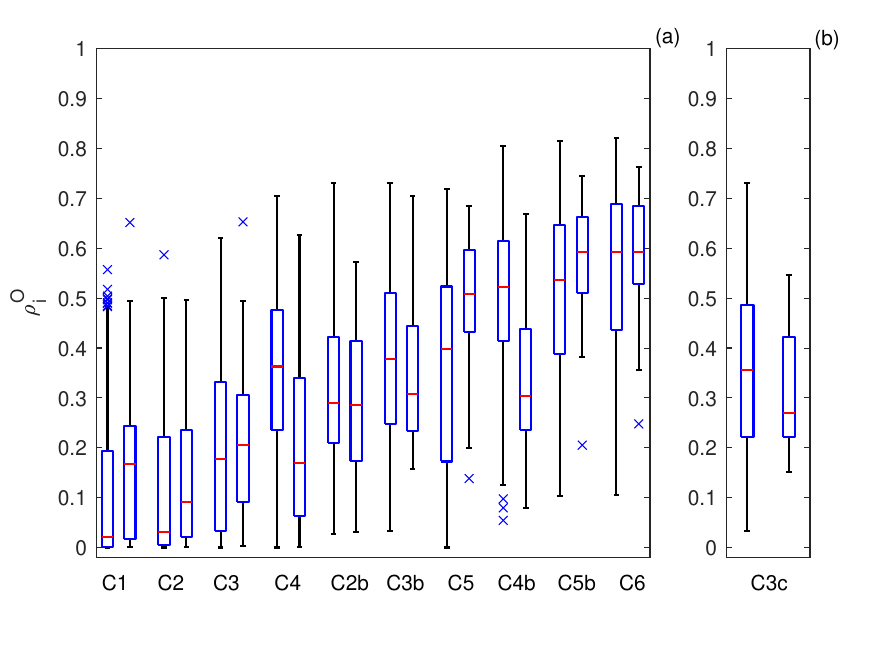}	
\caption{Boxplots of the values of $\rho_{i}^{O}$ for the stable non-hubs and stable hubs (left and right of each pair, respectively) for the ten described classifications are shown in (a) for increasing $N_{M}$. Panel (b) shows the results for a further classification, \textbf{C3c}, considered in the discussion section of this paper. The median is shown by a horizontal line in the box, extending from the lower to the upper quartile. Whiskers extend for 1.5 times the interquartile range or until the limit of the data is attained. Outliers are shown as crosses.}
\label{fig.part}
\end{figure}

Fig.~\ref{fig.part}a shows boxplots of the participation coefficient based on the out-degree $\rho_{i}^{O}$ (there was no qualitative difference if the $\rho_{i}^{I}$, i.e.~the equivalent term based on the in-degree were adopted instead) as a function of the ten classifications (each pair of boxplots) and if the vertices were classified as a stable non-hub (left-hand box) or stable hub (right-hand box) in Table \ref{table.ntypes}. For \textbf{C1} and \textbf{C2} virtually all the non-hub vertices are ultra-peripheral ($\rho_{i}^{O} \sim 0$) or peripheral nodes ($\rho_{i}^{O} \lesssim 0.625$), but for \textbf{C2b} onwards, the upper whisker for the non-hubs extends beyond $\rho_{i}^{O} = 0.625$, meaning that some nodes have a connector functionality. For \textbf{C1}-\textbf{C4} the great majority of hubs are of the provincial hub type ($\rho_{i}^{O} \sim 0.3$), but \textbf{C2b}-\textbf{C4b} are much more likely to include connector hubs ($0.3 < \rho_{i}^{O} \le 0.75$). The large number of modules in the classifications \textbf{C5b} and \textbf{C6} results in a dominance of connector hubs, with provincial hubs being very rare. Classifications \textbf{C4b}, \textbf{C5b} and \textbf{C6} have vertices approaching the kinless node categorisation ($\rho_{i}^{O} > 0.8$), while the kinless hub type was not observed for our network. Hence, classifications \textbf{C2b}-\textbf{C4b}, \textbf{C4} and \textbf{C5} feature a favourable diverse functionality of the nodes. In particular, \textbf{C3b} is preferable not only from the viewpoint of the modularity entropy (as in Fig.~\ref{fig.entropy}), but also for the diversity of the functionality of its nodes.

\subsection{Fluid Mechanical Properties of Stable Hubs and Non-Hubs}

\begin{figure}
\centering 
\includegraphics[width=0.8\textwidth, angle=0]{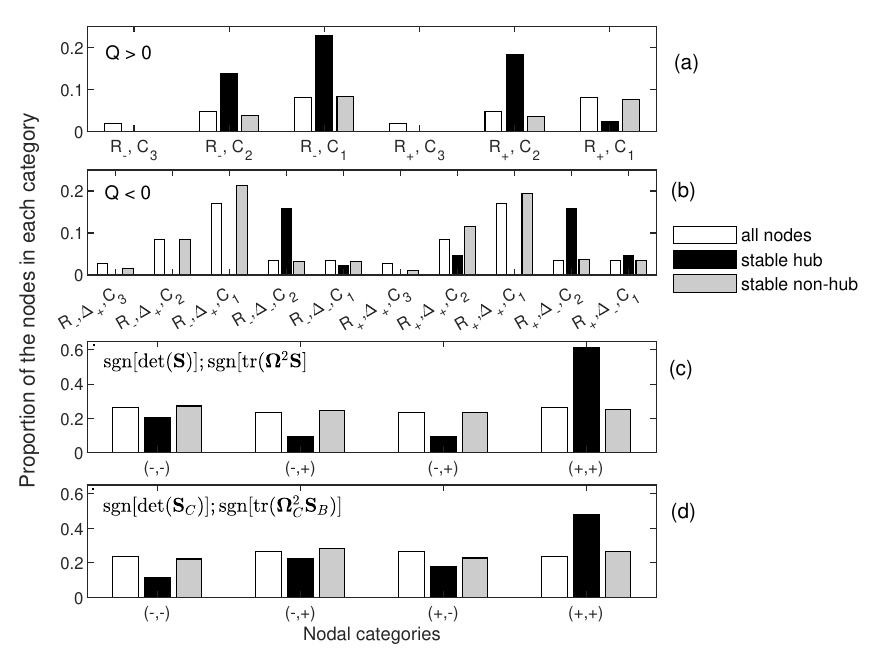}	
\caption{The proportion of all nodes (white), stable hubs (black) and stable non-hubs (grey) for different combinations of the signs of $Q$, $R$, and $\Delta$ indicated by $\pm$ subscripts, and the rank order of $\Vert\boldsymbol{\Omega}_{C}\Vert^{2}$ (abbreviated using ``C'') relative to the normal enstrophy and normal straining when considered in descending order of magnitude are given in panels (a) and (b). Hence, ``$C_{2}$'' means that $\Vert\boldsymbol{\Omega}_{C}\Vert^{2}$ is ranked second, and if $Q > 0$ then $\Vert\boldsymbol{\Omega}_{B}\Vert^{2}$ is ranked first. Panels (c) and (d) give the signs of the classic production terms and the non-normal production and interaction production, respectively. The order in which these terms are considered is indicated in the top-left of each panel and then the relevant sign combinations are provided on the horizontal axis.}
\label{fig.signs}
\end{figure}

Fig.~\ref{fig.signs} reports the features, in terms of the sign and ranking of the VGT invariants, for the stable hub nodes (black bars), stable non-hubs (grey bars) and for all the nodes (white bars). All the terms used in classifications \textbf{C1} to \textbf{C3b} feature in panels (a) and (b), while the signs of the production terms, which feature in the remaining classifications are given in panels (c) and (d). Note that for compactness we state the rank order of $\Vert\bm{\Omega}_{C}\Vert^{2}$ relative to $\Vert\bm{\Omega}_{B}\Vert^{2}$ and $\Vert\bm{S}_{B}\Vert^{2}$, rather than using $\xi$ and $\chi$ explicitly. 
Also, we do not give the signs of $-\mbox{det}\left(\bm{S}_{B}\right)$ and $\mbox{tr}\left(\bm{\Omega}_{B}^{2}\bm{S}_{B}\right)$ as these are opposite to one another for $\Delta > 0$, with the sign of the former given by the sign of $R$, which features in panels (a) and (b). 

Panel (a) focuses on the $Q > 0$ cases, and among these enstrophy-dominated nodes the stable non-hubs show very similar relative frequencies of VGT invariant sign combinations as compared to all nodes. In contrast, stable hub nodes have very distinct behaviour, which is not symmetrical in $R$, with strong non-normality when $R<0$ ($\mbox{rnk}\left(\Vert\bm{\Omega}_{C}\Vert^{2}\right) = 1$ being more frequent) and lower non-normality where $R>0$. This distinction highlights why classification \textbf{C3b} out-performs \textbf{C3} as it can capture that there are a greater number of hubs for $Q > 0, R < 0$ than $Q > 0, R > 0$ and also that those hubs arise where the non-normality dominates the dynamics.

Panel (b) shows that stable hubs rarely occur where $Q < 0$ and $\Delta > 0$, and not at all if, in addition, $R < 0$. Given the typical clockwise Lagrangian path around the $Q$-$R$ plane, $Q < 0$, $\Delta > 0$, $R < 0$ is where vorticity is first established (as the flow crosses the zero-discriminant line $\Delta = 0$ at $R < 0$, and the eigenvalues of $\bm{A}$ transition from all being real to one real value and a conjugate pair, resulting in $\Vert\bm{\Omega}_{B}\Vert^{2} > 0$).
This asymmetry in the VGT behaviour concerning $R$ explains why it is important for a classification to capture its sign.
Panel (b) also shows that the eigenvalues of $\bm{A}$ are real ($\Delta < 0$), stable hubs are preferentially associated with $\Vert\bm{\Omega}_{C}\Vert^{2} < \Vert\bm{S}_{B}\Vert^{2}$, i.e.~the $C_{2}$ state as the normal enstrophy is zero for $\Delta < 0$.
For a Gaussian velocity gradient, the PDF in the $\Delta<0$ region decays with $Q$ and is independent of $R$ \cite{Carbone2023}. Classification \textbf{C2} can capture the decay but misses the skewness, crucial for the energy cascade and characterizing flows even at very low Reynolds numbers \cite{carbone20,johnson20,Carbone2023}.
Indeed, this is sufficiently extreme in the $\Delta<0$, $R > 0$ region that synthetic statistical models that are more advanced than the Gaussian PDF, and capture the marginal behavior for $Q$ and $R$, cannot capture their joint behavior \cite{k17}.

These results highlight why neither \textbf{C1} nor \textbf{C2} is sufficient for effective classification of the VGT dynamics since different signs of $Q$, $R$ and $\Delta$ correspond to dynamically different configurations of the VGT. Let us highlight two examples of this in the following using the results in Fig.~\ref{fig.signs}:
Panel (a) shows that where $\Delta > 0$ and $\Vert\bm{\Omega}_{B}\Vert^{2} > \Vert\bm{S}_{B}\Vert^{2}$ the preference is for hubs to arise when non-normality is large. However, classification \textbf{C1} groups the $Q > 0$ cases from panel (a) with the $\Delta > 0$ cases in panel (b), thus conflating those configurations.
Panel (b) shows that for $R < 0$, stable hubs only arise significantly where normal enstrophy is zero and non-normality is small relative to normal straining (the $R_{-}, \Delta_{-}, C_{2}$ case). No stable hubs occur where normal enstrophy is non-zero but smaller than the normal straining irrespective of the non-normality (the $R_{-}, \Delta_{+}$ cases for any of  $C_{1}$, $C_{2}$, or $C_{3}$). However, classification \textbf{C2} conflates the $\Delta_{+}$ and $\Delta_{-}$ cases shown in panel (b).

While the occurrence of stable hubs displays marked asymmetries concerning the signs of the VGT invariants (particularly the sign of $R$), stable non-hubs show a more symmetric behavior. Panel (b) in Fig.~\ref{fig.signs} indicates there is a preferred state for the stable non-hubs, which is approximately symmetric in $R$, and arises where $\Delta > 0$ and $\Vert\boldsymbol{\Omega}_{C}\Vert^{2}$ is the largest magnitude term. Hence, in two of the six regions defined by \textbf{C3} where hubs arise only rarely (and only at all for $R > 0$), non-hubs occur preferentially. Relative to the flow in general, $Q < 0$, $\Delta_{+}$, $C_{1}$ for either sign of $R$ is the only state that leads to a preferential occurrence of non-hubs.

The last two panels in Fig.~\ref{fig.signs}, (c) and (d), focus on the signs of the production terms. Given that the signs of the normal enstrophy and normal straining are opposite (where the normal enstrophy is non-zero) with the sign of the normal straining equal to the sign of $R$ \cite{k18}, we focus on the classical strain and enstrophy production in (c) and the non-normal and interaction production in (d). The results on these panels are correlated due to the decompositions (\ref{eq.Qterms}) and (\ref{eq.Rterms}), and show that stable hubs arise preferentially when all production terms are positive, while the occurrence of stable non-hubs is independent of the signs of the production terms (the grey bars and white bars are similar throughout these panels).
Thus, for these stable hubs, $-\mbox{det}\left(\bm{S}_{C}\right) > 0$ and $\mbox{tr}\left(\bm{\Omega}_{C}^{2}\bm{S}_{B}\right) > 0$ is sufficient to give $-\mbox{det}\left(\bm{S}\right) > 0$ and $\mbox{tr}\left(\bm{\Omega}^{2}\bm{S}\right) > 0$ even though, depending on the sign of $R$, one of the normal enstrophy production or the normal strain production must be negative. The easiest way for this situation to arise is with $R > 0$, imposing $-\mbox{det}\left(\bm{S}_{B}\right) > 0$, and $\Delta < 0$, which gives $\mbox{tr}\left(\bm{\Omega}_{B}^{2}\bm{S}_{B}\right) = 0$. Panel (b) shows that in this situation, stable hubs arise where $\Vert\bm{\Omega}_{C}\Vert^{2} < \Vert\bm{S}_{B}\Vert^{2}$, i.e.~$R_{+}$, $\Delta_{-}$, $C_{2}$. Hence, the stable hubs with all positive production terms are concentrated on the right Vieillefosse tail \cite{vieillefosse82} and with minimal impact from the non-normality (these hubs are interconnecting flow states that are dominated by normal straining behaviour \cite{betchov56,vieillefosse84,maurizio20,k24}).   

\begin{figure}
\centering 
\includegraphics[width=0.8\textwidth, angle=0]{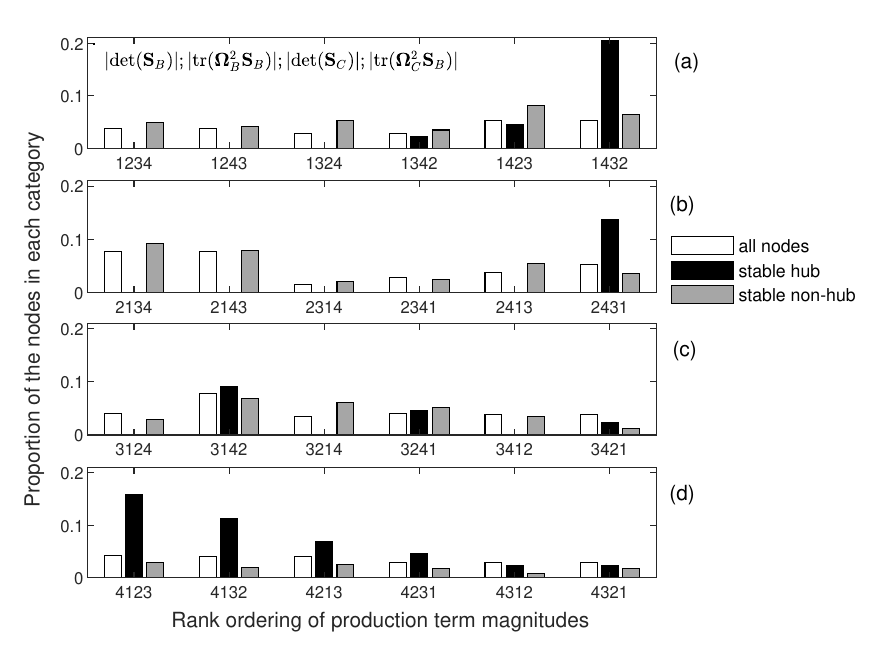}	
\caption{The proportion of all nodes (white), stable hubs (black) and stable non-hubs (grey) for each of the twenty-four permutations of the rank order of the four production terms. This order is stated beneath each set of three bars (ranking in descending order), with the order of the four terms stated in the top panel.} 
\label{fig.prod}
\end{figure}

By design, the part of the definition of the distinct nodes that did not feature in any of the classifications was the relative magnitude of the four production terms on the right-hand sides of (\ref{eq.Rterms}), to avoid the number of communities/modules tending to the number of nodes. Fig.~\ref{fig.entropy} indicates that entropy ``saturation'' is arising by $N_{M} = 64$. However, given that the magnitude of the production terms does not feature as part of the classifications, and is therefore not directly dependent upon them, it is interesting to determine if the stable hubs and non-stable hubs exhibit clear differences concerning this aspect of the flow. This is illustrated in Fig.~\ref{fig.prod} where all twenty-four possible rankings of the magnitudes of the four production terms are shown, with each panel featuring the six permutations given that the normal strain production is ranked, respectively, first (a), second (b), third (c) and fourth (d). While the stable non-hubs (grey bars) exhibit broadly similar characteristics to all nodes (white bars), the stable hubs (black bars) strongly differ from the average behavior of all nodes. In particular, there is one instance in each of panels (a) and (b), and two instances in (d) where there is a very strong excess probability of a stable hub arising. In panels (a) and (b), stable hubs are characterized by the normal enstrophy production magnitude, $|\mbox{tr}(\boldsymbol{\Omega}_{B}^{2}\mathbf{S}_{B})|$, being the smallest term. This configuration shows up most readily where $\Delta < 0$, all the eigenvalues of the VGT are real and there is no local enstrophy. In both (a) and (b), $|\mbox{det}(\bm{S}_{C})|$ remains the second smallest term and the relative standing of $|\mbox{det}(\bm{S}_{B})|$ and $|\mbox{tr}(\bm{\Omega}_{C}^{2}\bm{S}_{B})|$ changes from first to second between panels (a) and (b). The instance on the second row is of particular interest since one might assume that $|\mbox{det}(\mathbf{S}_{B})| > |\mbox{tr}(\bm{\Omega}_{C}^{2}\bm{S}_{B})|$ when $|\mbox{det}(\bm{S}_{B})| > |\mbox{det}(\bm{S}_{C})|$. However, this is not the case, highlighting the crucial role played by the alignment between the normal straining eigenvectors and the non-normal vorticity vector for these stable hubs. 
The stable hubs in Fig.~\ref{fig.prod}d arise when the normal strain production is the smallest in magnitude and the normal enstrophy production is the largest. This configuration occurs when $Q > 0$ and the normal straining is not vanishing, as that might lead to $\vert\mbox{det}(\bm{S}_{C})\vert > \vert\mbox{tr}(\bm{\Omega}_{B}^{2}\bm{S}_{B})\vert$.
For the stable hubs to be more common in the left-most case (`4123') rather than the second case (`4132') it should be the case that the normal straining must be closely aligned with the normal vorticity compared to the non-normal vorticity. It was shown by \cite{k18} (their Figure 17) that it is only where $Q < 0, \Delta > 0, R > 0$ that there is a preferential alignment between the eigenvectors of $\bm{S}_{B}$ and $\bm{\omega}_{C}$, and this is at a $45^{\circ}$ angle. However, we know from Fig.~\ref{fig.signs}b that stable hubs rarely occur in this region (there are no stable hubs for $R_{+}$, $\Delta_{+}$, $C_{1}$ and only a small number for $R_{+}$, $\Delta_{+}$, $C_{2}$). In contrast, where $Q > 0$, \cite{k18} found a very strong alignment between the eigenvector of either the most compressive (where $R > 0$) or most extensive (where $R < 0$) eigenvalue of $\bm{S}_{B}$ and the normal vorticity, $\bm{\omega}_{B}$. Hence, the stable hubs in Fig.~\ref{fig.prod}d preferentially exhibit this strong alignment of the normal strain rate and vorticity and are associated with vortical regions.


\section{Discussion}
The modularity entropy and participation coefficient reported in Fig.~\ref{fig.entropy} and \ref{fig.part} suggests that the best classifier of the VGT Lagrangian dynamics is \textbf{C3b}. Hence, beyond the sign of the principal invariants $Q,R$ and the discriminant, $\Delta$, we do not need to explicitly consider the production terms featured in (\ref{eq.Delta}), in contrast to the expansion of the $Q$-$R$ plane by \cite{luthi09}. However, because both $Q$ and $R$ and, thus, $\Delta$ may be written entirely in terms of the eigenvalues of $\bm{A}$, the tensor non-normality is excluded from consideration in classifications \textbf{C1}--\textbf{C3}. Thus, including $\Vert\bm{\Omega}_{C}\Vert^{2}$ to give \textbf{C3b} results in a significantly improved classification of the VGT dynamics.

\begin{figure}
\centering 
\includegraphics[width=0.8\textwidth, angle=0]{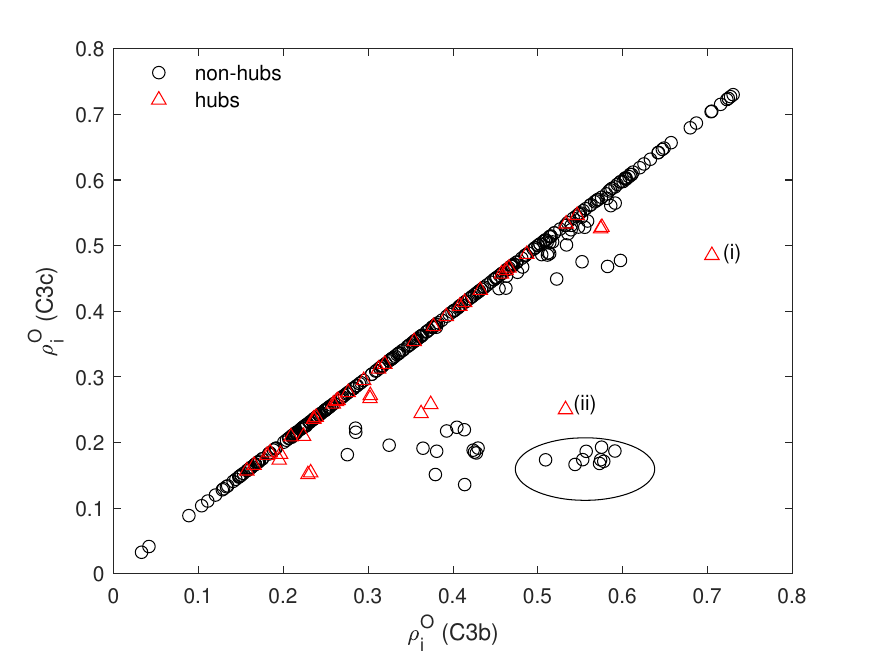}
\caption{Values of the participation $\rho^{O}_{i}$ for classifications \textbf{C3b} and \textbf{C3c} with hubs shown as triangles and non-hubs as circles. The labels highlight two individual hubs and the ellipse a group of nine non-hubs that are discussed further in the text.} 
\label{fig.rho3bc}
\end{figure}

However, Fig.~\ref{fig.signs}b suggests that in terms of nodal function, there is no clear distinction between $R < 0, \Delta < 0$ and $R > 0, \Delta < 0$, with stable hubs occurring at an approximately equal frequency in these two regions and, in both instances with a similar elevated frequency where $\Vert\bm{\Omega}_{C}\Vert^{2} < \Vert\bm{S}_{B}\Vert^{2}$. This raises the possibility that \textbf{C3b} can be enhanced by simplifying from six to five regions in the $Q$-$R$ plane by merging the regions where $\Delta < 0$ for either sign of $R$ (and then retaining differences in $\xi$ and $\chi$). This gives classification \textbf{C3c} with $N_{M} = 14$ communities rather than $N_{M} = 16$. The modularity entropy of this new case is $\mathcal{E} = 2.832$ compared to $\mathcal{E} = 2.975$ for \textbf{C3b}. In addition, inserting this value into (\ref{eq.grad}) gives $d\mathcal{E}/d N_{M} = 0.212$, which is less than that for \textbf{C3b} (where $d\mathcal{E}/d N_{M} = 0.228$). Furthermore, we have included \textbf{C3c} in panel (b) of Fig.~\ref{fig.part} and there is no obvious change in nodal functionality in terms of the boxplots for $\rho_{i}^{O}$ for the stable hubs and non-hubs. This would suggest that our original \textbf{C3b} classifier is more effective. This is borne out in Fig.~\ref{fig.rho3bc}, which shows that for all nodes where there is a significant change, \textbf{C3c} induces a loss of nodal diversity as stable hubs or non-hubs with a tendency to act as connectors in \textbf{C3b} are peripheral under \textbf{C3c}. This can be seen by looking more closely at the two hubs that undergo the greatest change, which are labelled (i) and (ii) in the figure. These are very similar in terms of the fluid dynamics they represent: Both are where $\Delta < 0$, $R < 0$, with positive signs for $-\mbox{det}(\bm{S})$, $\mbox{tr}(\bm{\Omega}^{2}\mathbf{S})$, $-\mbox{det}(\bm{S}_{C})$ and $\mbox{tr}(\bm{\Omega}_{C}^{2}\mathbf{S}_{B})$, and with the interaction production, $|\mbox{tr}(\bm{\Omega}_{C}^{2}\mathbf{S}_{B})|$, the largest magnitude of the production terms. 
A key distinguishing characteristic is the proportion of the out-degree to flow states with different signs for $\Delta$, $R$ or $Q$ (rather than to flow states with the same signs). The two highlighted stable hub nodes had approximately 50\% of the out-degree to the same $\Delta < 0$, $R < 0$ cases, with 25\% to $\Delta < 0$, $R > 0$ and 25\% to $Q < 0, \Delta > 0$, $R < 0$ vertices. This is in marked contrast to the other two stable hubs in the $\Delta < 0$, $R < 0$ region, which had 88\% of their out-degree to the $\Delta < 0$, $R < 0$ cases, with 4\% to $\Delta < 0$, $R > 0$ and 8\% to $Q < 0, \Delta > 0$, $R < 0$. In other words, the change from \textbf{C3b} to \textbf{C3c} no longer contrasts those hubs that are key to the flow state transitions across the left Vieillefosse tail, gaining the flow local enstrophy and, thus, vortical properties.


\section{Conclusion}
Adopting an entropy form for the network modularity (\ref{eq.entropy}) and combining this with a consideration of nodal participation (\ref{eq.part})-(\ref{eq.z}), allows us to identify an effective classification for the Lagrangian dynamics of the velocity gradient tensor (VGT).
Such a classification, here denoted as \textbf{C3b},
first combines two well-known approaches, given as \textbf{C1} and \textbf{C2}, based on the signs of the VGT principal invariant, $R$, and discriminant, $\Delta$, as well as on the signs of the second and third invariants, $Q$ and $R$, respectively. Then, rather than adopting this classification (denoted as \textbf{C3}), it is beneficial to incorporate the non-normality of the VGT into the classification.
The most effective direction to take this non-normality into account is not that pursued by \cite{luthi09}, who implicitly added the combined effect of non-normal production and interaction production to the normal strain production and normal enstrophy production (\ref{eq.Rterms}). Instead, our results show it is more effective to establish the relative standing of $\Vert\bm{\Omega}_{C} \Vert^{2}$ compared to the two quantities that define $Q$, namely $\Vert\bm{\Omega}_{B} \Vert^{2}$ and $\Vert\bm{S}_{B} \Vert^{2}$ (\ref{eq.Qterms}), which we accomplish using the two quantities, $\xi$ and $\chi$, defined in (\ref{eq.xichi}).
Hence, the most effective classification according to the criteria presented here, consists of a three-dimensional space. Two axes and six regions are defined in the $Q$-$R$ plane, where all the terms can be written using the VGT eigenvalues, while the third axis incorporates information not captured by the VGT eigenvalues. This is made explicit using a Schur decomposition of the VGT, as in (\ref{eq.schur}) and (\ref{eq.Qterms}).  
Including the VGT non-normality turned out to be key to obtaining an effective classification, and this is because it reflects crucial aspects of the VGT dynamics.
Following e.g., \cite{vieillefosse82,cantwell92,ooi99}, several studies have produced maps for the VGT dynamics that reflect the three distinct contributions in (\ref{eq.evolveQ}) and (\ref{eq.evolveR}): the restricted Euler equations for $Q$ and $R$, the viscous term and the part of the dynamics due to the deviatoric part of the pressure Hessian \cite{vieillefosse82,cantwell92,chevillard08,zhou15,Tom2021}. Such studies have shown that the diffusive effect of the viscous term is similar everywhere, pulling the flow towards the origin on the $Q$-$R$ plane, while the deviatoric pressure Hessian introduces intricate dynamics, that is the focus of several modelling efforts \cite{chevillard08,wilczek14,johnson16,carbone20,Yang2024}.
The inclusion of $\Vert\bm{\Omega}_{C} \Vert^{2}$ provides a means to capture these non-local influences on the tensor in our discrete classification of the VGT dynamics. Such influences introduce torques, which are reflected in the non-normal part of the VGT.
The resulting classification \textbf{C3b} differs from others based purely on the VGT topology \cite{PC87} or on the production terms \cite{luthi09}, and it reflects the role of particular nodes that provide a means for the VGT to transition across different regions on the $Q$-$R$ plane \cite{k24}.


Applying methods from complex networks theory \cite{newman04,guimera05} to networks encoding the VGT dynamics can shed light on its intricate time evolution and preferential configurations, for example, by extracting communities of VGT states and studying their interaction. Namely, different nodes within a community can act as hubs or non-hubs, and those hubs or non-hubs occur when certain fluid dynamical constraints are met. This one-to-one correspondence between the network and the flow configurations constitutes a promising research direction. In general, using a discrete network that allows for a comprehensive description of the VGT through a limited number of degrees of freedom is a promising approach to better understand the VGT dynamics, for practical flow modelling and development of flow control methods.

\section*{Acknowledgements}
The authors gratefully acknowledge the Leverhulme Trust International Fellowship 2023-014 that permitted CK's visit to Bayreuth and enabled
this research.










\end{document}